# Non-linear vortex dynamics in the mixed state of superconducting *a*-MoGe and NbN thin films using low-frequency two-coil mutual inductance technique

Somak Basistha[1], Soumyajit Mandal, John Jesudasan, Vivas Bagwe and Pratap Raychaudhuri

*Tata Institute of Fundamental Research, Homi Bhabha Road, Mumbai 400005, India*

## Abstract

We use a two-coil mutual inductance technique to investigate the non-linear response of the vortex lattice of two type-II superconducting thin films, namely a very weakly pinned amorphous Molybdenum Germanium (*a*-MoGe) and a moderate-to-strongly pinned Niobium Nitride (NbN). We observe a strong dependence of the magnetic shielding response of the superconductors on the ac excitation amplitude in the primary coil of the two-coil setup. The sample response is studied through the evolution of the radial profile of the induced current density with increasing drive amplitude, which gets significantly modified by the effect of flux creep. We develop a computation scheme where we iteratively combine the coupled Maxwell-London equations for the geometry of the two coils and the sample involved, with a model developed by Coffey and Clem, to analyze the non-linear ac response. The central result of this analysis is that the effect of flux creep gives rise to a strong nonlinearity in the electrodynamic response in the vortex state of the superconducting thin films, that extends down to very low amplitudes of ac excitation. Our results also show that at subcritical low frequency ac drives, the vortex viscosity is exponentially larger than the Bardeen-Stephen estimate. We present a simple scheme to obtain the intrinsic value of the pinning force constant, which otherwise gets affected due to flux creep even at very low ac drives and point out some outstanding issues that need to be addressed in future theoretical and experimental studies.

[1] Email: somak.b.94@gmail.com

## 1. Introduction

When a type II superconductor is subjected to a magnetic field greater than its lower critical field $H_{c1}$, the field penetrates inside the superconductor in the form of quantised flux tubes each carrying a flux of $\phi_0 = \frac{h}{2e}$ Wb, called vortices[1]. When an external current is passed through the superconductor, each vortex experiences a Lorentz force, $F_L = \phi_0 \boldsymbol{J} \times \hat{\boldsymbol{n}}$, where $\boldsymbol{J}$ is the current density and $\hat{\boldsymbol{n}}$ the unit vector along the magnetic field. In a perfectly clean superconductor, this results in the movement of the vortices, which gives rise to dissipation. This dissipation gives rise to an effective viscous force, $-\eta\dot{\boldsymbol{u}}$, where $\boldsymbol{u}$ is the position of the vortex and $\eta$ is the viscous drag coefficient which can be estimated from the upper critical field, $H_{c2}$[1] and the normal state resistivity, $\rho_N$, using the Bardeen-Stephen model[2] as, $\eta = \frac{\mu_0 H_{c2}\phi_0}{\rho_N}$. $\eta$ can be recast as $\eta = \frac{B\phi_0}{\rho_{ff}}$ where $B$ is the applied dc magnetic field and $\rho_{ff}$ is the resistivity in the flux flow regime[2] of the superconductor, where the velocity of vortices, across the superconductor, is only limited by the viscous drag. In real superconductors, defects in the solid act as potential wells that pin the vortices, thereby altering this behaviour. In the presence of a dc current, dissipation occurs only above a finite critical current density[3] ($J_c$), at which the Lorentz force ($F_L$) can overcome the pinning force. The situation is more complex under an oscillatory drive, like an ac current or ac magnetic field. Here each vortex can undergo small oscillatory motion within the pinning potential even when $|\boldsymbol{J}| < J_c$, leading to dissipation. The dynamics of vortices is determined by the interplay of the viscous force and the restoring pinning force. Understanding the dynamics of vortices is therefore of paramount importance both from a fundamental standpoint and from the aspect of technological applications[4,5,6,7,8,9].

The two-coil mutual inductance technique[10,11,12,13,14] is a simple, yet elegant method to investigate the low-frequency electrodynamics of superconducting films, in frequency range of tens of $kHz$. In this technique, the superconducting thin film is sandwiched between a primary and a secondary coil such that the film partially shields the magnetic field produced by the primary coil, from the secondary. The electrodynamic response of the superconductor is obtained from the complex mutual inductance ($M = M' + iM''$) between the two coils. This technique has widely been used to measure the penetration depth in superconducting films and to study the Berezinski-Kosterlitz-Thouless transition in ultrathin films[15]. Recently[16], this technique was extended to study the electrodynamic response in the vortex state of NbN and $a$-MoGe superconducting films under a low frequency subcritical ac drive (tens of $kHz$ frequencies). These results showed that for a subcritical $kHz$ drive the vortex viscosity $\eta$ exceeds the Bardeen-Stephen estimate by several orders of magnitude. Although there are some theories predicting such an enhancement of vortex viscosity, like vortex-pin (Vinokur *et al.* [17]), vortex-vortex interactions (Carruzzo *et al.* [18]) and more recently, inelastic relaxation of quasiparticles in the vortex core (Pashinsky *et al.* [19]) the precise origin of this large enhancement is still debated.

Two-coil measurements are typically performed with very small ac excitations where the response of the superconductor is assumed to be in the linear regime. This assumption is valid typically when the induced current density, $J$, in the superconductor is much smaller than the critical current density $J_c$, for example in zero field for most superconductors. On the other hand, in the presence of vortices, processes such thermally activated flux flow (TAFF) and flux creep (FC) can significantly reduce the range of the ac excitation amplitude over which the superconductor is in the regime of linear response. Therefore, it is extremely important to understand the non-linear response of the superconductor to interpret the data. Previous studies on the low frequency ac non-linear response of vortex dynamics using ac susceptibility measurements[20,21,22,23,24], have focussed on the hysteretic regime of the superconductor[25,26]. On the other hand, these studies did not investigate of the non-linear ac response in the non-hysteretic regime and its implications on the vortex parameters pertaining to the non-linear ac response.

In this paper, we investigate the non-linear electrodynamic response in the vortex state of two superconductors, namely, a $20\ nm$ thick $a$-MoGe thin film with very weak pinning and a moderate-to-strongly pinned NbN thin film of thickness $5\ nm$. In both samples, we observe a strong dependence of $M$ on the amplitude of the ac drive in the primary coil that persist much above the magnetic field where the hysteresis loop closes. We develop a computation scheme to analyze the data in the non-linear regime and observe that the effect of flux creep plays a major role in the non-linear response of the vortex state, even at very low amplitudes of the ac drive, in the non-hysteretic regime.

## 2. Theoretical background

Vortices in type-II superconductors, under small oscillatory excitations, behave like forced-damped harmonic oscillators. They are influenced by three forces: (i) the viscous drag force, (ii) the restoring force due to the combined effect of pinning and inter-vortex interactions and (iii) the Lorentz force due to the oscillatory current. Gittleman and Rosenblum's[27] (GR) model describes vortex dynamics for a single vortex, neglecting vortex mass term and thermal effects. Under these approximations the equation of motion of the vortex is given by:

$$\eta\dot{\boldsymbol{u}} + \alpha\boldsymbol{u} = \phi_0 \boldsymbol{J}^{\boldsymbol{ac}} \times \widehat{\boldsymbol{n}} \qquad (1)$$

$\boldsymbol{J}^{\boldsymbol{ac}}$ is the ac drive current density, $\widehat{\boldsymbol{n}}$ is the unit vector along the magnetic field of the vortex, $\eta$ is the viscous drag coefficient of the vortices in absence of pinning and flux creep, and $\alpha$ is the restoring force constant (Labusch parameter[28]) of the vortex. It is a mean-field model[29], which captures the vortex dynamics if the thermally activated motion of the vortices is slow compared to the excitation frequency. For a harmonic solution ($\sim e^{i\omega t}$) of eqn. (1), we get $\boldsymbol{u} = \phi_0 \frac{\boldsymbol{J^{ac}} \times \hat{\boldsymbol{n}}}{(\alpha + i\omega\eta)}$. This value when substituted into the London equation[30] we get[29]:

$$\boldsymbol{A} = -\mu_0 \lambda_L^2 \boldsymbol{J^{ac}} + \boldsymbol{u} \times \boldsymbol{B} = -\mu_0 \left( \lambda_L^2 + \frac{\phi_0 B}{\mu_0(\alpha + i\omega\eta)} \right) \boldsymbol{J^{ac}} = -\mu_0 \tilde{\lambda}^2 \boldsymbol{J^{ac}} \qquad (2)$$

Eqn. (2) assumes a form similar to the usual London equation, where the London penetration depth[31] ($\lambda_L$) is replaced by the effective complex penetration depth, $\tilde{\lambda}$. The term $\frac{\phi_0 B}{\mu_0(\alpha + i\omega\eta)}$ captures the effect of the pinning force and viscous drag, and in the low frequency limit, $\omega \ll \alpha/\eta$, reduces to the square of the well-known Campbell penetration depth[32,33,34] given by, $\lambda_C = \left( \frac{\phi_0 B}{\mu_0 \alpha} \right)^{1/2}$. It is often useful to rewrite $\tilde{\lambda}^2$ in terms of the complex vortex resistivity, $\rho_v$:

$$\tilde{\lambda}^2 = \lambda_L^2 + \frac{\phi_0 B}{\mu_0(\alpha + i\omega\eta)} = \lambda_L^2 + \frac{\lambda_C^2}{(1 + i\omega\tau_0)} = \lambda_L^2 - \frac{i\,\rho_v}{\mu_0 \omega} \qquad (3)$$

where $\tau_0 = \eta/\alpha$ is the vortex relaxation time and $\rho_v$ is expressed in terms of dc flux flow resistivity ($\rho_{ff}$) as[29,30]: $\rho_v = \rho_{ff} \frac{i\omega\tau_0}{1 + i\omega\tau_0}$; $\rho_{ff} = \frac{B\phi_0}{\eta} = \frac{B}{B_{c2}} \rho_n$, where $\rho_n$ is the normal state resistivity.

Eqn. (2) does not take thermal effects into consideration. At finite temperatures, the strength of the pinning barriers being finite, there is a finite probability of the vortices to hop from one pinning site to another neighbouring pinning site, owing to their thermal energy. In the absence of any external drive, these hops have no preferential direction and hence there is no net motion of the vortices. This process, called thermally activated flux flow (TAFF), has the effect of relaxing the restoring pinning force over characteristic time scale of vortex hop. This effect becomes particularly important when measurements are done at low frequencies. We use a model proposed by Coffey and Clem[35] (CC model), who added a random Langevin force, dependent on the effective pinning potential barrier ($U$), to simulate the thermal motion and then solve it similar to that of a particle undergoing Brownian motion in a periodic potential[36,37]. This leads to a modification of eqn. (3), where the complex resistivity takes the form[38],

$$\rho_v \rightarrow \rho_v^{TAFF} = \rho_{ff}\frac{\epsilon + i\omega\tau}{1 + i\omega\tau} \qquad (4)$$

where $\tau$ is the relaxation rate and $\epsilon$ is a dimensionless parameter which is a measure of the weight of the TAFF phenomenon. They are given as:

$$\epsilon = \frac{1}{I_0^2(\nu)} \qquad (5\text{ (a)})$$

$$\tau = \tau_0\frac{I_0^2(\nu)-1}{I_1(\nu)I_0(\nu)} \qquad (5\text{ (b)})$$

where $\nu = \frac{U}{2k_BT}$ and $I_0$ and $I_1$ are the zeroth and first order modified Bessel functions of the first kind. Consequently, CC model has three temperature dependent parameters, $\alpha, \eta$ and $U$. Furthermore, the normal skin depth ($\delta_{nf}$) arising from the electrodynamic response of the normal electrons introduces an additional correction such that eqn. (3) is modified into:

$$\tilde{\lambda}^2 = \left(\lambda_L^2 - i\frac{\rho_v^{TAFF}}{\mu_0\omega}\right)\Big/\left(1 + 2i\frac{\lambda_L^2}{\delta_{nf}^2}\right) \qquad (6)$$

Here, $\delta_{nf}$ has the phenomenological variation of the form $\sim\left(\frac{\left(\frac{2\rho_n}{\mu_0\omega}\right)}{1-f(t,h)}\right)^{1/2}$ which is complementary to the variation of $\lambda_L$, where $\lambda_L^2(t,h) = \frac{\lambda_L^2}{f(t,h)}$, both of which comes from analysing a superconductor in the framework of the two-fluid model. Here $\rho_n$ is the normal state resistivity and $f(t,h) = (1-t^4)(1-h)$[35,30] with $t = T/T_c$ and $h = H/H_{c2}(t)$. Since, $\lambda_L \sim 361\ nm$ for NbN and $588\ nm$ for $a$-MoGe[16], $\frac{2\lambda_L^2}{\delta_{nf}^2}$ << 1 and hence we can drop the this term from the denominator of eqn. (6).

Apart from TAFF, there is another motion that the vortices exhibit, called flux creep[30]. When an external drive in the form of a current density ($\boldsymbol{J}$) is present, the Lorentz force density ($\sim \boldsymbol{J} \times \boldsymbol{B}$) acts on the vortices and the original pinning potential becomes a tilted washboard model, which introduces a difference in the rate of forward (in the direction of the Lorentz force) hopping and backward (against the direction of the Lorentz force) hopping. Flux creep has the effect of reducing $U$. Kim and Anderson[39,40] first worked out this phenomenon, where the effective pinning potential gets attenuated by the external drive and is given by: $U(J) \sim U_0 \left(1 - \frac{|J|}{J_{c0}}\right)$, where $J_{c0}$ is the critical current density without flux creep. However later works by Beasley[41] and more recent works by Griessen[42] and Lairson[43] showed that this expression gets modified to:

$$U(J) \sim U_0 \left(1 - \frac{|J|}{J_{c0}}\right)^{3/2} \qquad (7)$$

when the anharmonicity of the pinning potential is taken into consideration. In the regime of $J \ll J_{c0}$, an inverse power law barrier (Feigel'man *et al.*[44,45]):

$$U(J) \sim U_c \frac{\left[\left(\frac{J_{c0}}{|J|}\right)^{\mu} - 1\right]}{\mu} \qquad (8)$$

and a logarithmic barrier (Zeldov *et al.* [46,47])

$$U(J) \sim U_c ln\left(\frac{J_{c0}}{|J|}\right) \qquad (9)$$

has been used to study flux creep in high temperature superconductors, mostly in the regime of collective creep in vortex solids.

Apart from the reduction in $U$, the effect of flux creep also results in $J$ dependence of $\alpha$ and $\eta$. Prozorov *et al.*[48] showed that, considering a cubic anharmonicity of the pinning potential, $\alpha$ gets modified as

$$\alpha(J) \sim \alpha_0 \left(1 - \frac{|J|}{J_{c0}}\right)^{1/2} \qquad (10)$$

Within the tilted washboard model this functional form is derived from the local curvature of the pinning potential at the new equilibrium position of the vortex in the presence of a current density. On the other hand, the $J$ dependence of vortex viscosity $\eta$ comes from the $J$ dependence of $U$. It has been suggested that $\eta$ gets enhanced due to vortex-vortex and vortex-pin interactions. Vinokur *et al.*[17] and Carruzzo *et al.*[18] introduced an empirical form of this enhanced $\eta$, which is given by:

$$\eta(J) \sim \eta_{BS} e^{U(J)/k_B T} \quad (11)$$

where $\eta_{BS}$ is the BS viscosity. Qualitatively, this functional form predicts that $\eta$ will exceed the Bardeen-Stephen value at subcritical current and reach the Bardeen-Stephen limiting value for $J \geq J_c$.

The final form of the CC equation that we use is:

$$\tilde{\lambda}^{-2} = \lambda^{-2} + i\delta^{-2} = \left(\lambda_L^2 - i\frac{\rho_{ff}}{\mu_0 \omega}\frac{\epsilon + i\omega\tau}{1 + i\omega\tau}\right)^{-1}$$

$$= \left(\lambda_L^2 - \frac{\rho_{ff}}{\mu_0 \omega\,(1 + (\omega\tau)^2)}\left(\omega\tau(\epsilon - 1) + i(\epsilon + (\omega\tau)^2)\right)\right)^{-1} \qquad (12)$$

Here, $\lambda$ accounts for the inductive response, while $\delta$ measures the dissipative response of the superconductor. In this work, we use eqn. (12) along with eqns. (7), (10) and (11) to address the non-linear response of the vortex state, which arises from the $J$ dependence of $U$, $\alpha$ and $\eta$.

## 3. Sample and Experimental Details

*Sample details:* The samples under investigation are a $20\ nm$ thick superconducting $a$-MoGe thin film and a $5\ nm$ thick superconducting NbN thin film, both grown on a (100) MgO substrate, using pulsed laser deposition and dc magnetron sputtering respectively. Details of sample growth and characterisation can be found in refs. [49], [50] and [51]. The $T_c$ determined from the two-coil mutual inductance response (fig. (1)) is 7 $K$ for $a$-MoGe film and 12 $K$ for NbN film respectively. Furthermore, for both films $M'$ shows a sharp drop whereas $M''$ shows a narrow single peak just below $T_c$, confirming the homogeneous nature of our films.

The contrasting pinning properties of the two samples can be seen from the inset of fig.

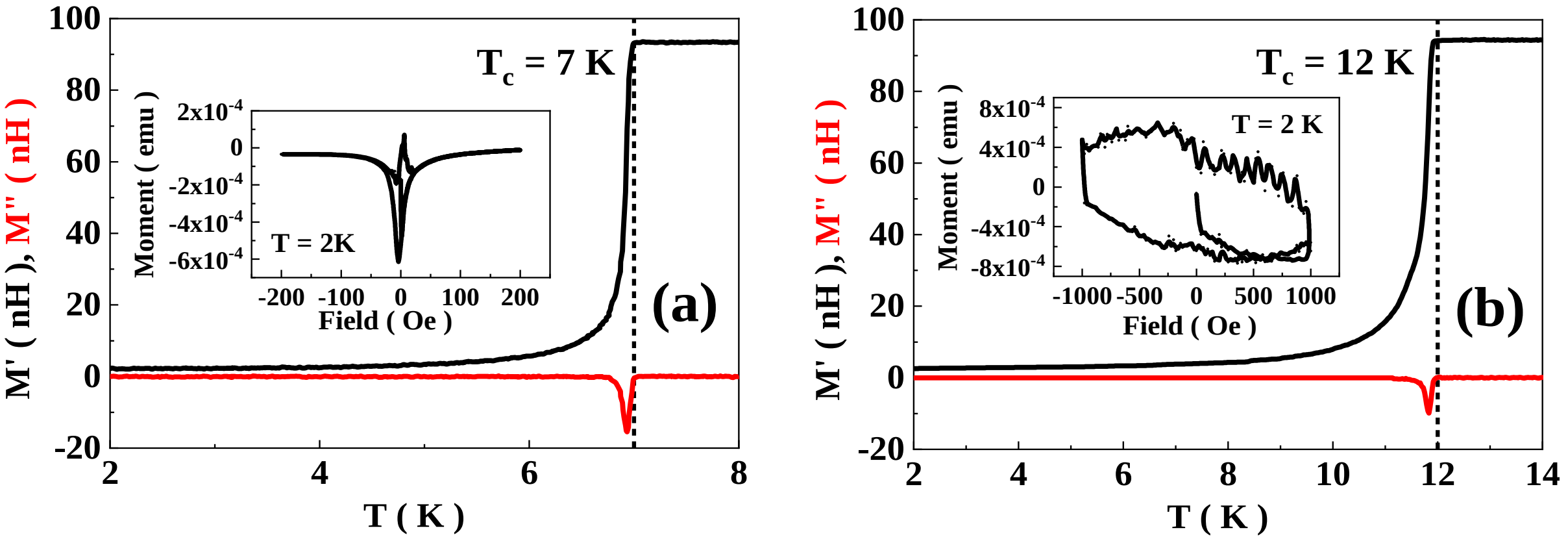

**Figure 1.** Zero-field temperature variation of $M'$ and $M''$ for **(a)** 20 $nm$ thick, superconducting $a$-MoGe film, and **(b)** 5 $nm$ thick, superconducting NbN film, both grown on (100) oriented MgO substrates. $M'$ and $M''$ are plotted as black and red solid lines respectively. The dashed lines correspond to $T_c$ which is ~ 7 $K$ for $a$-MoGe and ~ 12 $K$ for NbN. These measurements are performed with $I_{ac}$ ~ 0.05 $mA$ and $f = 30\ kHz$. The insets show variation of magnetic moment as a function of DC magnetic field $(m - H)$ loops for the same samples, performed using a SQUID-VSM at $T = 2\ K$.

(1 (a)) and (1 (b)) where we plot the magnetisation $(m)$ as a function of an applied DC magnetic field $(H)$ for the two films at 2 $K$, measured using a SQUID-vibration sample magnetometer (SQUID-VSM). Both samples show hysteresis in $m - H$ typical of a type II superconductor. However, while for NbN the hysteresis loop is considerably open at 1 $kOe$, for $a$-MoGe the loop closes around 25 $Oe$ showing its extremely weakly pinned nature. Therefore, in terms of pinning strength these two samples lie on two ends of the spectrum: the NbN film is strongly pinned while the pinning in $a$-MoGe thin film is extremely weak.

*Measurement of the low frequency electrodynamic response:* We probe the vortex low frequency dynamics, by measuring the ac shielding response of the superconductor, through measurement of the mutual inductance ($M$) between the two coils. In this setup[10,11,12], we sandwich an 8 $mm$ diameter superconducting thin film, grown using a shadow mask (bottom left inset of fig. (2)) in between a miniature quadrupolar primary coil and a dipolar secondary coil. Both the primary and secondary coils are wound on bobbins of 2 $mm$ diameter made from Delrin. An ac current with amplitude $I_{ac}$ and frequency, $f$ (= 30 $kHz$), is passed through the primary coil and the resulting in-phase ($V_{in}$) and out-of-phase voltage ($V_{out}$), in the secondary is measured using a lock-in amplifier. The complex mutual inductance between the two coils is given by $M'(M'') = V_{out}\ (V_{in})\ /\ (2\pi f I_{ac})$, where $M'$ corresponds to the inductive response and $M''$ corresponds to the dissipative response.

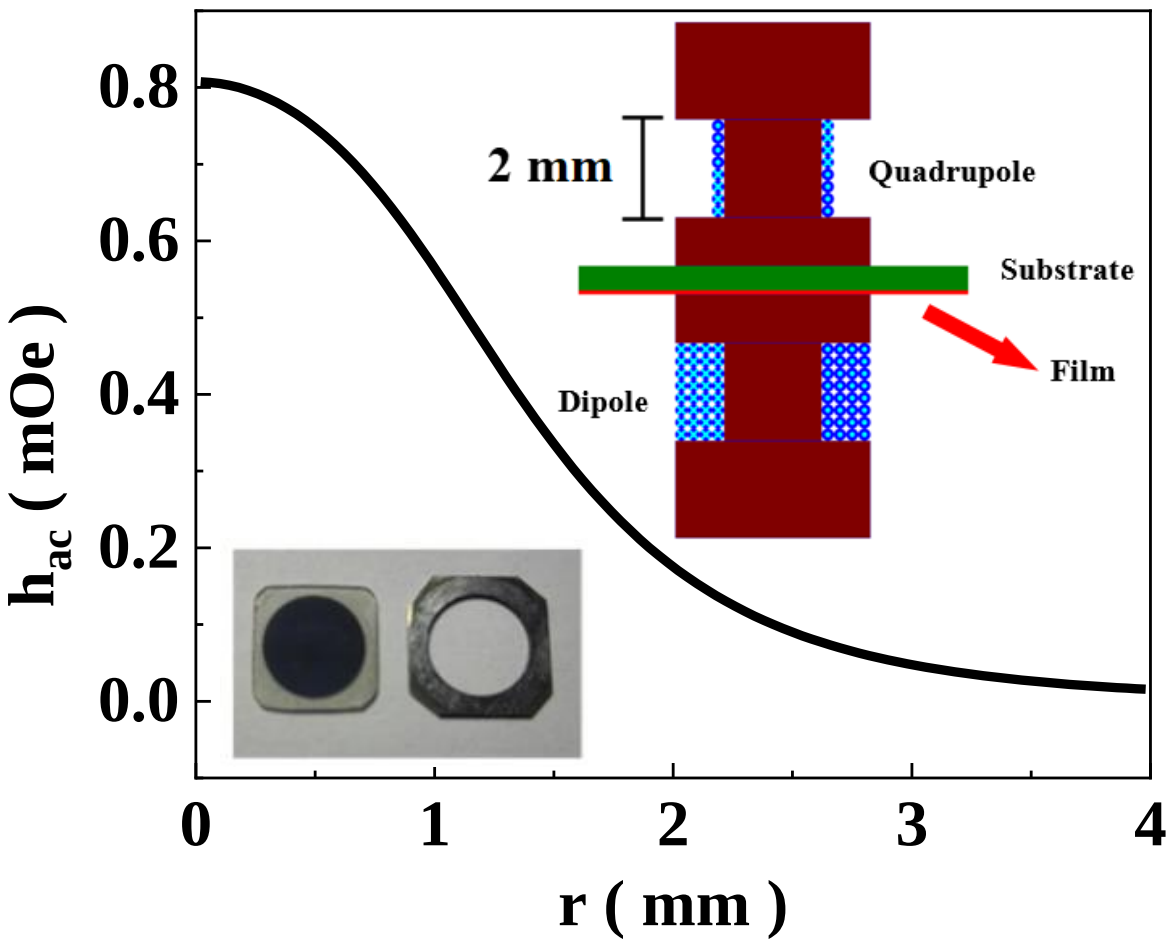

**Figure 2.** Radial variation of ac field amplitude ($I_{ac}$) generated across the sample plane from passing 0.05 $mA$ ac excitation through primary quadrupole coil (black); (inset: top right) schematic of the two-coil setup: quadrupole as primary coil (top) and dipole as secondary coil (bottom) with sample sandwiched in between. The coil wire diameter ( 50 $\mu m$ ) is drawn bigger than the actual for clarity; (inset: bottom left) 8 $mm$ diameter sample grown on MgO substrate (left) using stainless steel mask (right).

The schematic of the two-coil setup is given in the top right inset of fig. (2). The quadrupolar primary coil consists of 15 turns clockwise in one half of the coil and another 15 turns counter-clockwise in the other half. The use of a quadrupolar coil ensures fast radial decay of the ac magnetic field on the film, which minimises the effect of edge effects[52] and geometric[53,54] barriers. The ac magnetic field ($h_{ac}$) generated by the primary coil has a maximum ($h_p$) at the centre of the sample and decreases radially, reaching a value 0.02 $h_p$ at the edge of the film. For our coil geometry, $h_p/I_{ac}$, is 16 $Oe/A$ such that at the maximum drive amplitude of $15\,mA$, $h_p \sim 240\,mOe$. The dipolar secondary coil consists of $120$ turns wound in four layers. Both the coils are made of $50\,\mu m$ diameter copper wires. We study the amplitude dependent shielding response of the superconducting films at a fixed temperature of $2\,K$, at an operating frequency of $30\,kHz$ and at different applied dc magnetic fields. We ramp $I_{ac}$, from $0.05\,mA$ till $6\,mA$ (for $a$-MoGe) and $15\,mA$ (for NbN). The lower limit of $(I_{ac})$ is kept at $0.05\,mA$, as the signal level becomes comparable to the noise threshold below that value. The upper limit of the ac amplitude is chosen based on the magnitude, where $M'$ nears saturation or heating sets in the sample, whichever occurs earlier.

## 4. Results

In fig. (3), we summarize the ac shielding response of the two superconductors, as a function of the amplitude of ac excitation in the primary coil, at a temperature of $2\,K$ and at different applied dc magnetic fields. The peak of the ac magnetic field, generated on the superconductors, due to passage of the ac excitation is given in the top x-axis of each subplot of fig. (3). We observe with increasing $I_{ac}$, $M'$ (fig. (3 (a)) and (c)) increases monotonically towards the normal state value ($94\,nH$). $M''$ (fig. (3 (b)) and (d)) shows a dissipative peak, which becomes broader with increasing dc magnetic field. For $a$-MoGe we restrict the lower value of dc magnetic field to $100\,Oe$, to stay away from the hysteretic regime as indicated by SQUID measurements. For the same reason, for NbN, the dc magnetic field is restricted down to $2\,kOe$. A contrasting feature between $a$-MoGe and NbN, is that, in NbN, at low fields ($2\,kOe$ – $20\,kOe$), $M'$($M''$) varies weakly till about $I_{ac} \sim 2\,mA$ and then, a rapid variation sets in. For $a$-MoGe, there is a steady variation in $M'$ and $M''$ throughout the entire range of $I_{ac}$. This difference is owing to the much stronger pinning in the NbN film as compared to $a$-MoGe as we will show later.

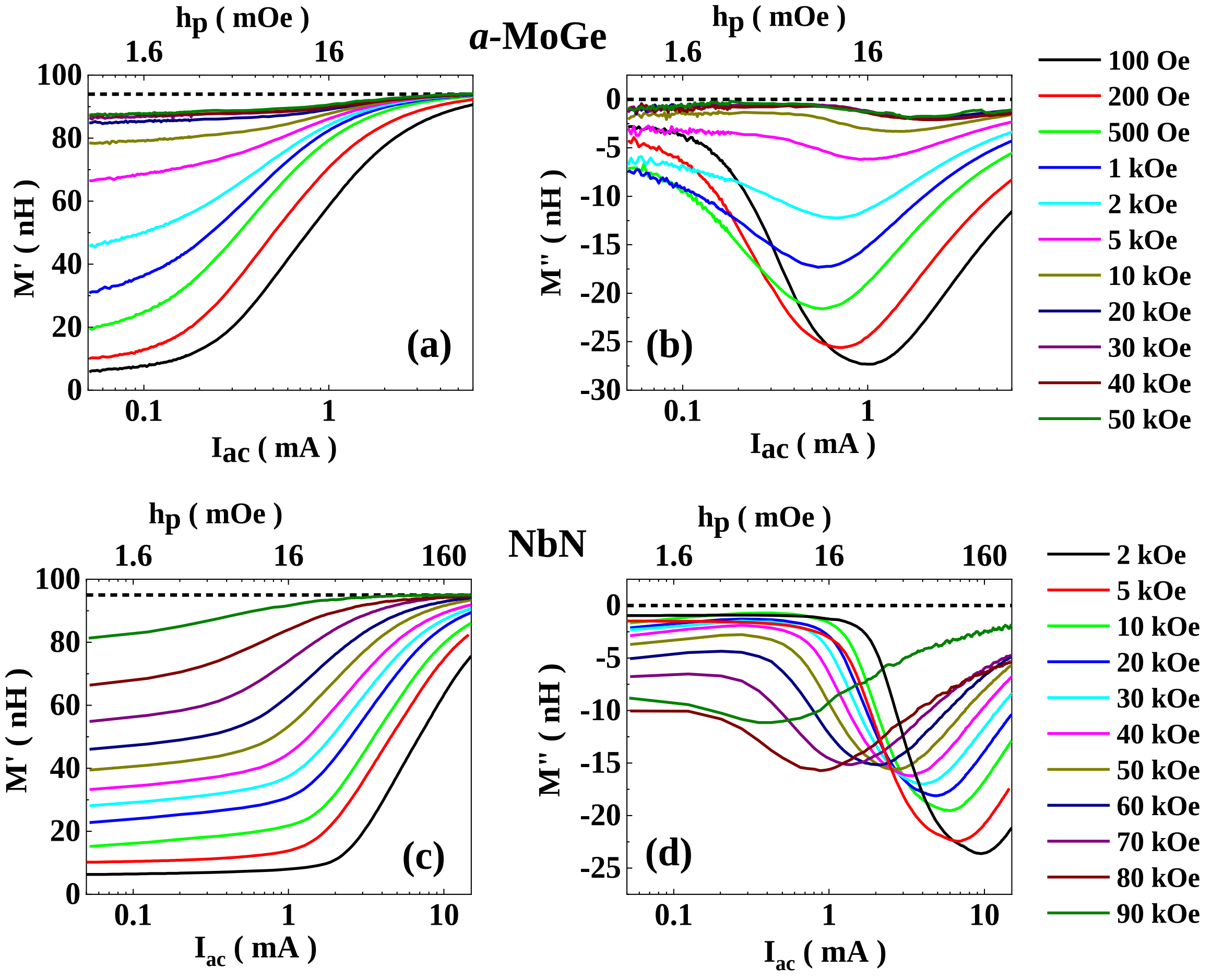

**Figure 3.** Magnetic field variation of $M'$ and $M''$ with $I_{ac}$ (bottom x-axis) and the corresponding $h_p$ (top x-axis) produced by the primary coil, for **(a) – (b)** $a$-MoGe and **(c) – (d)** NbN. The dotted lines in black show the $M'$ and $M''$ value in the normal state, which is 94 $nH$ for $M'$ and 0 $nH$ for $M''$. The measurements are done at 2 $K$ and at a frequency of 30 $kHz$.

## 5. Discussions

### 5.1 Effect of ac excitation amplitude on the electromagnetic shielding response of a superconductor

To understand the effect of $I_{ac}$ on the shielding response of a superconducting film, we start by analysing the induced current density in the films due to the magnetic field produced by the primary coil. When $I_{ac}$ is very low, the magnitude of the induced current density ($J_{ac}$) on the superconducting film, is much smaller than $J_c$. Therefore, the effect of flux creep on $U$, $\alpha$ and $\eta$ are negligible and $\tilde{\lambda}$ can be taken as a constant. In this regime, we can calculate $J_{ac}$ by numerically solving the coupled Maxwell and London equations using a procedure developed by Turneaure *et al.*[10,11]. Fig. (4 (a)) and (b) shows typical profiles of the real and imaginary part of $J_{ac}$ ($J'_{ac}$ and $J''_{ac}$ respectively).

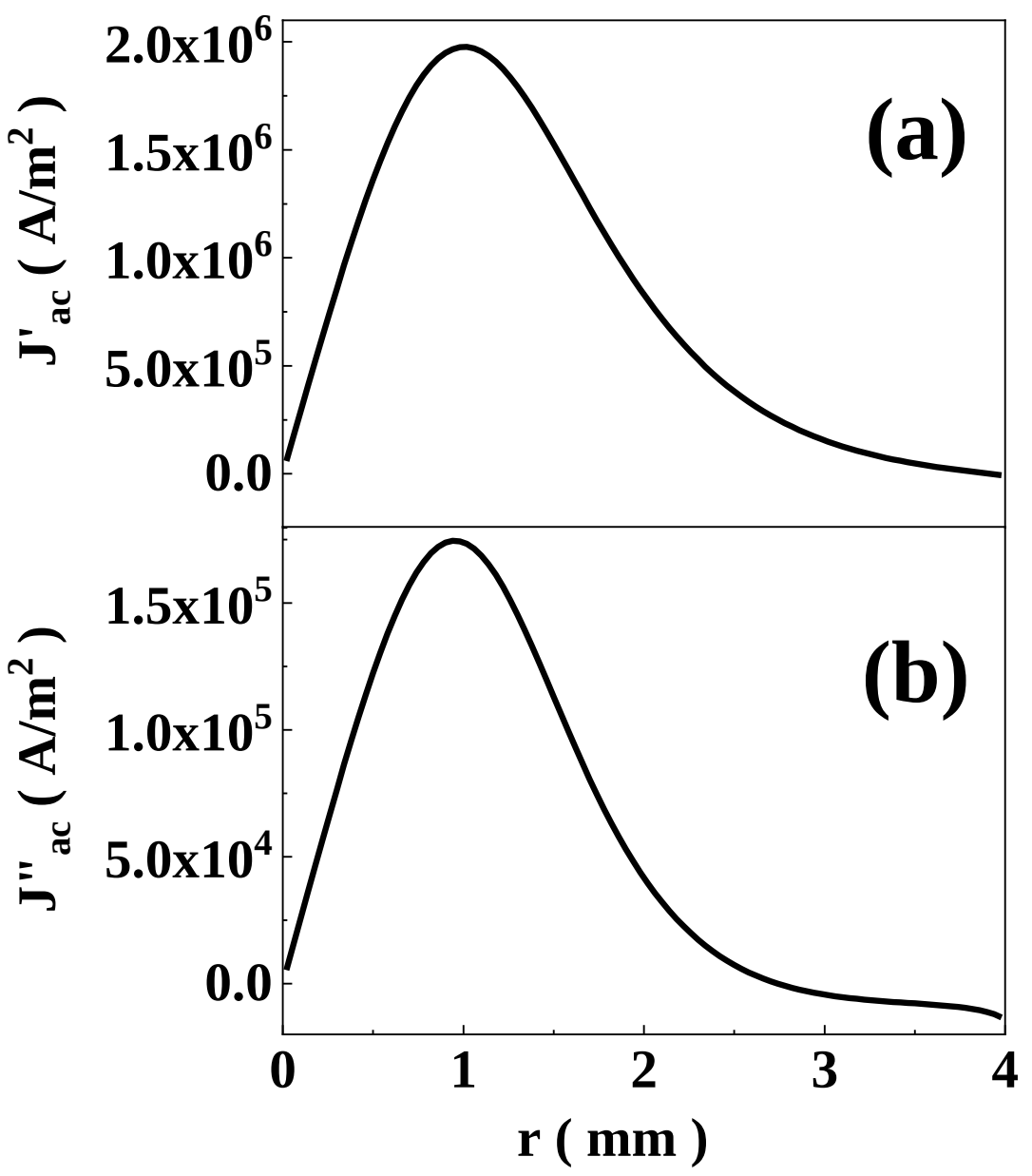

**Figure 4. (a) – (b)** show the radial profile of the real $(J'_{ac})$ and imaginary $(J''_{ac})$ components of the induced current density $(J_{ac})$ at $I_{ac} = 0.05\,mA$ without considering the effect of flux creep. We have taken $\lambda = 1890\,nm$ and $\delta = 6.6 \times 10^6\,nm$. These values correspond to $\alpha_0 \sim 46\,Nm^{-2}$, $\frac{U_0}{k_B} \sim 13.2\,K$ and $\eta = \eta_{BS} e^{U_0/k_B T}$ where $\eta_{BS} \sim 1.47 \times 10^{-8}\,Nsm^{-2}$ at $T = 2\,K$ and $B = 1\,kOe$.

When the effect of creep is included, $\tilde{\lambda}$ gets locally modified based on the local value of $J_{ac}$, and therefore $\tilde{\lambda}(J_{ac})$ acquires a radial dependence. This, in turn modifies the radial profile of $J_{ac}$. To obtain a consistent description of the modified $\tilde{\lambda}$ and $J_{ac}$ due to flux creep for a given value of $U_0, \alpha_0$ and $J_c$, we adopt an iterative procedure. We start by solving the coupled Maxwell-London with a constant $\tilde{\lambda}$, corresponding to the lowest $I_{ac}$, and obtain a profile for $J_{ac}$. Next, we use these $J_{ac}$ values, to account for the effect of flux creep on $\tilde{\lambda}$, using eqns. (7), (10), (11) and (12) and obtain $\tilde{\lambda}(r)$. We then use this modified $\tilde{\lambda}(r)$ to recalculate $J_{ac}$. This procedure is repeated, till a convergence in $J_{ac}$ is achieved. The details of the iterative procedure are given in Appendix A.

Fig. (5) shows the radial variation of $J'_{ac}$ and $J''_{ac}$, corresponding to a constant $\tilde{\lambda}$, and that obtained after convergence of the iterative procedure discussed above, at four different values of $I_{ac}$ for $a$-MoGe at 1 $kOe$ and 2 $K$. As expected, with increasing $I_{ac}$, the effect of flux creep becomes more prominent. In the profile of $J'_{ac}$ (fig. (5 (a)), (c), (e) and (g)), the peak around $1\ mm$ from the sample centre starts to flatten out and at large drive current, there appears a slight dip. This profile can be qualitatively understood from the fact that close to the radius where the bare $J'_{ac}$ is large, the effect of creep is to cause a steep increase in $\lambda$; this acts as negative feedback eventually decreasing $J'_{ac}$. (These profiles are similar to the ones previously calculated in ref. [55], [56] using a slightly different method). $J''_{ac}$ (fig. (5 (b)), (d), (f) and (h)) increases steadily with $I_{ac}$, and at larger $I_{ac}$ the dissipative response dominates.

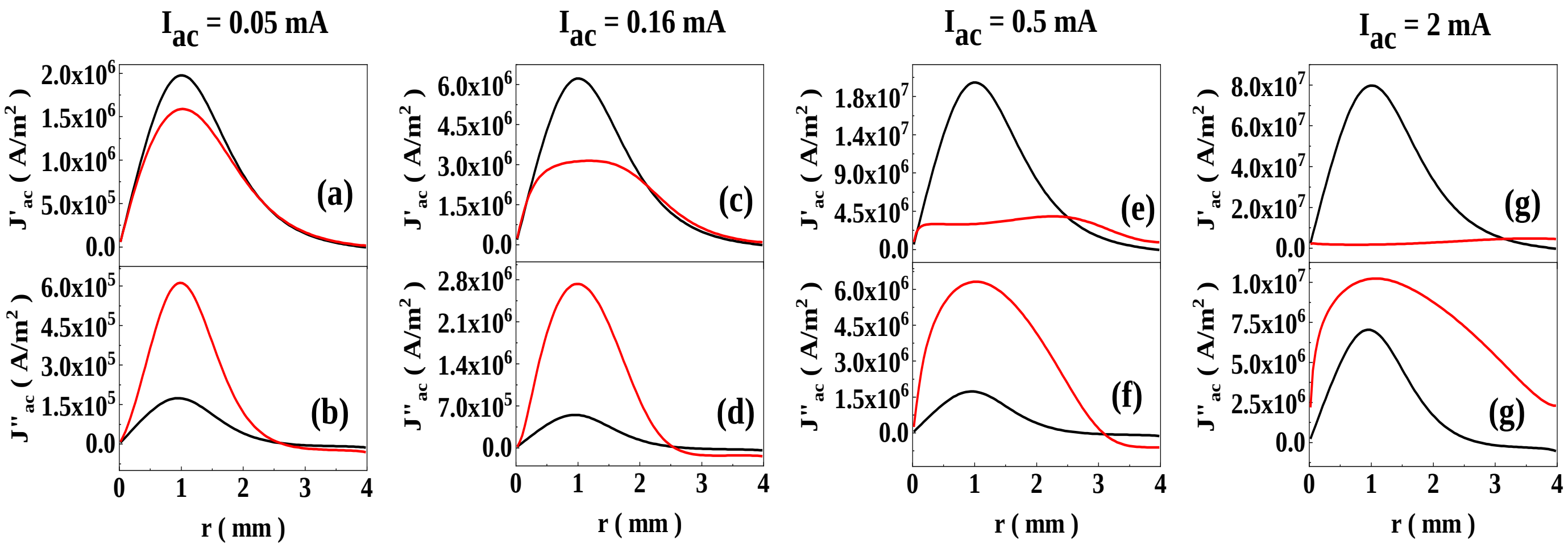

**Figure 5.** Radial profile of the real ($J'_{ac}$) and imaginary ($J''_{ac}$) components of $J_{ac}$ without (black solid lines) and with (red solid lines) the effect of flux creep for **(a) – (b)** $I_{ac} = 0.05\ mA$ , **(c) – (d)** $I_{ac} = 0.16\ mA$ , **(e) – (f)** $I_{ac} = 0.5\ mA$ and **(g) – (h)** $I_{ac} = 2\ mA$ . We have taken $\alpha_0 \sim 46\ Nm^{-2}, \frac{U_0}{k_B} \sim 13.2\ K$ and $J_c \sim 4.8 \times 10^7 Am^{-2}$ along with $\eta = \eta_{BS} e^{U_0/k_B T}$ where $\eta_{BS} \sim 1.47 \times 10^{-8}\ Nsm^{-2}$ at $T\ =\ 2\ K$ and $B\ = 1\ kOe$.

To calculate $M'(M'')$ and fit the data, we calculate the combined magnetic field arising from the calculated current profile in the film and the magnetic field produced by the primary coil to find the induced voltage in the secondary coil[10,11,12]. $\alpha_0, U_0$ and $J_c$ are used as adjustable parameters to obtain the best fit to the $M'\ vs\ I_{ac}$ curve. The evolution of $M'(M'')$ with $I_{ac}$ for $a$-MoGe at $1\ kOe$ and $2\ K$ calculated in this way along with the experimental data is shown in fig. (6). We note that this computation technique is successful in qualitatively capturing the variation of $M'$ over the entire range. Quantitatively, the variation of $M'$ is very well captured from low to intermediate $I_{ac}$, but a small deviation is observed at higher $I_{ac}$, indicating a breakdown of the perturbative expansions based on which, formulae in eqns. (7), (10) and (11) are derived. On the other hand, for $M''$, we capture the qualitative trend, but the calculated value is larger than the measured value. As we will discuss later, that although this analysis largely captures the quantitative variation of $M'$, there are some serious quantitative disagreements in some current regimes.

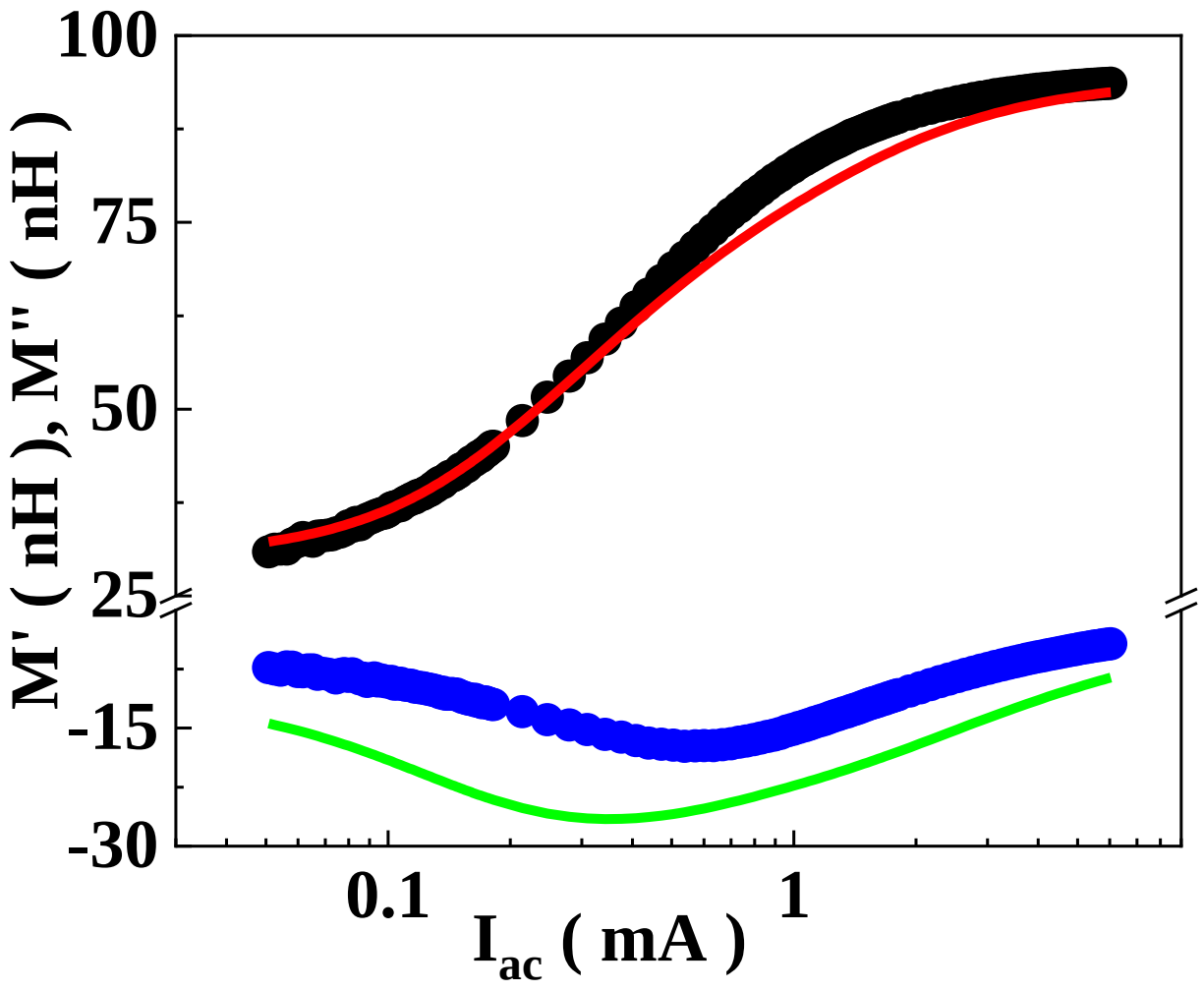

**Figure 6**. show the AC amplitude dependence of $M'$ and $M''$ for $a$-MoGe at $2\,K$, $1\,kOe$. The black and blue solid circles represent the experimental data for $M'$ and $M''$ respectively, while the red and green solid lines represent the fitted response to $M'$ and $M''$. The fit parameters are: $\alpha_0 \sim 46\ Nm^{-2}$, $\frac{U_0}{k_B} \sim 13.2\ K$ and $J_c \sim 4.8 \times 10^7 Am^{-2}$ along with $\eta = \eta_{BS} e^{U_0/k_B T}$ where $\eta_{BS} \sim 1.47 \times 10^{-8}\ Nsm^{-2}$.

## 5.2 Fits to the magnetic field variation of the electromagnetic shielding response of *a*-MoGe and NbN

*a-MoGe*: Fig. (7 (a) – (f)) show the fits to the amplitude dependence of $M'$ and $M''$ for $a$-MoGe thin film at different applied DC magnetic fields. We observe that the variation of $M'$ with $I_{ac}$ is quantitatively captured, with a small deviation at higher drive amplitudes, while the trend in $M''$ is captured qualitatively, albeit with an overestimation. The fits to $M'$ gets better with increasing dc magnetic field. We have used the Beasley model of flux creep in our fitting routine. This model has been used previously while analysing $I - V$ characteristics of a similar film by Buchachek *et al.*[57]. The fits to the complete dataset for $a$-MoGe is provided in Appendix

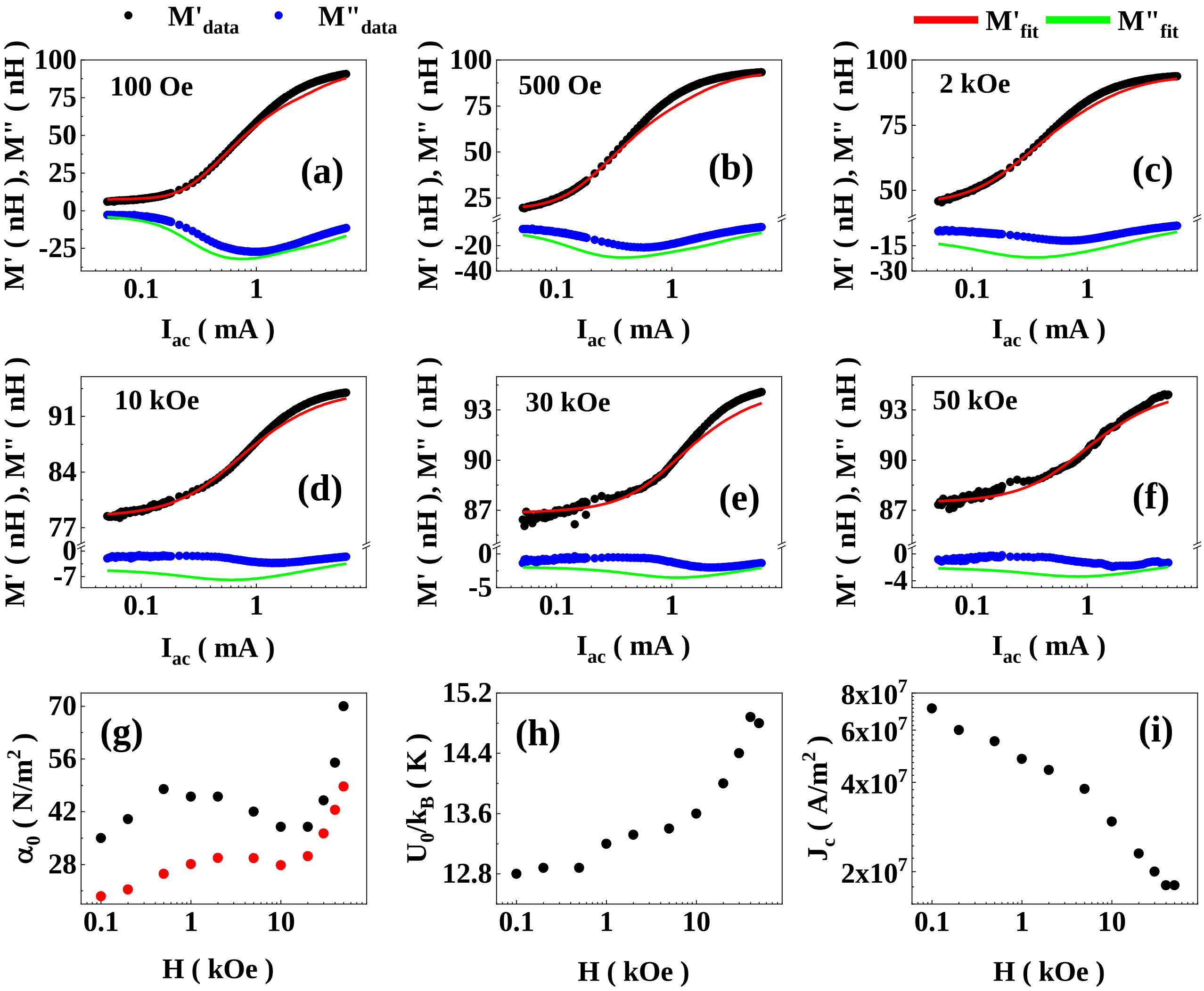

**Figure 7. (a) – (f)** $I_{ac}$ dependence of $M'$ and $M''$ of 20 $nm$ $a$-MoGe at different applied DC magnetic fields. The black and blue solid circles are the experimentally measured $M'$ and $M''$ respectively while the red and green solid lines are the fits to $M'$ and $M''$. Magnetic field dependence of **(g)** $\alpha_0$ (ref. [16], using GR model, neglecting flux creep, (red solid circles) and this analysis (black solid circles)), **(h)** $\frac{U_0}{k_B}$ and **(i)** $J_c$ which are the parameters used for fitting with the Beasley model of flux creep at $T = 2\ K$ and $f = 30\ kHz$.

B. The variation of $\alpha_0, U_0,$ and $J_c$ with magnetic field is shown in fig. (7 (g) – (i)). The values of $J_c$ are of the same order of magnitudes as the ones obtained earlier from dc current voltage $(I - V)$ measurements on a similar sample[58]. The variation of $\alpha_0$ follows the same trend as the one obtained from the value of $\tilde{\lambda}$ calculated from $M$ measured at the lowest $I_{ac}$ using the analysis based on the GR model (eqn. (3)) which neglects the effect of flux creep. This is similar to the analysis done in ref. [16]. However, the absolute values are larger by a factor of $1.5-2$. We will discuss the origin of this difference later. The magnitude of $U_0$ also differs from ref. [16], but this is not surprising since in ref. [16] at low temperatures, $U_0$ was calculated by extrapolating its value close to $T_c$, thereby introducing large errors.

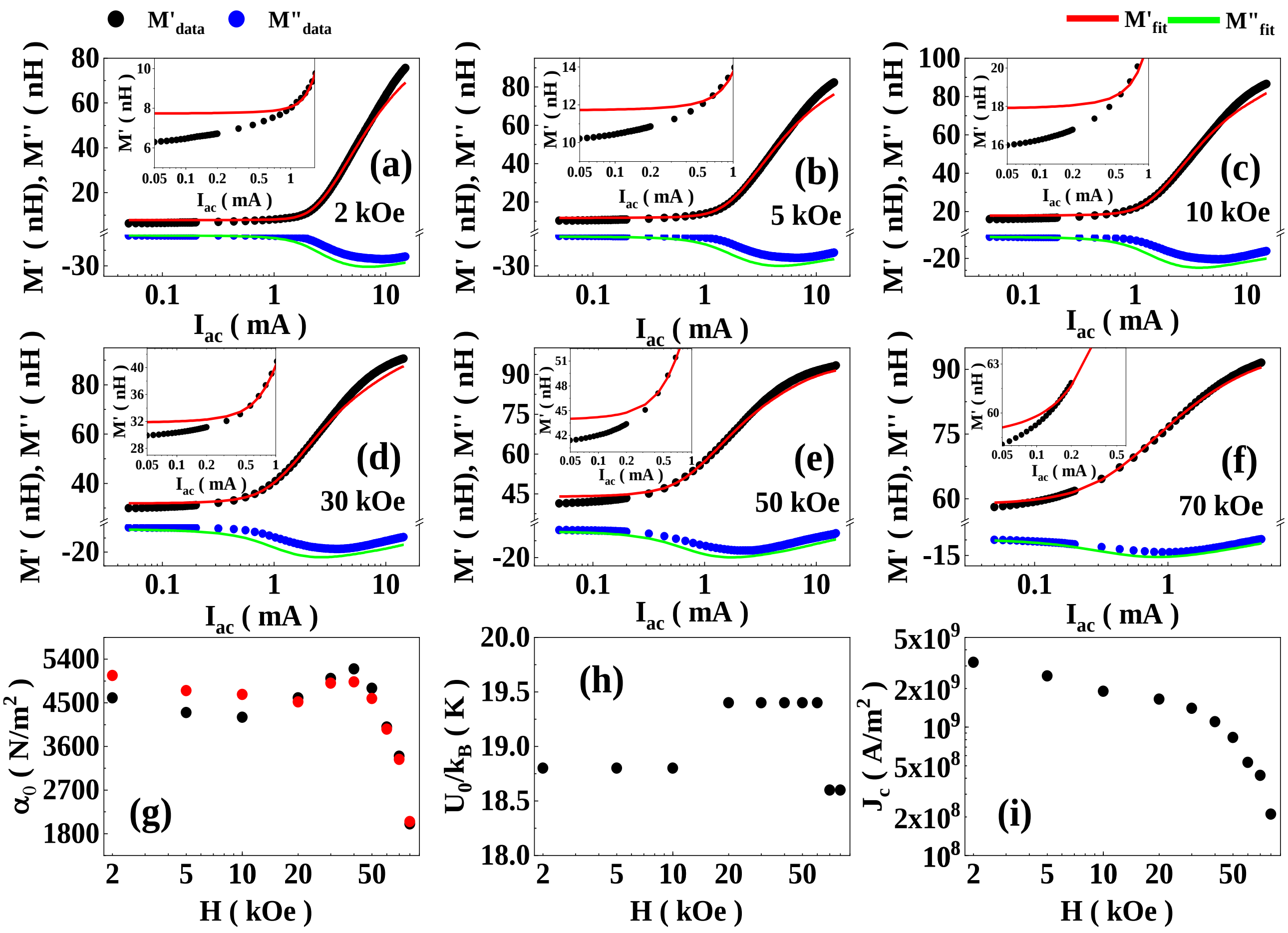

**Figure 8. (a) – (f)** $I_{ac}$ dependence of $M'$ and $M''$ of 5 $nm$ NbN at different applied DC magnetic fields. The black and blue solid circles are the experimentally measured $M'$ and $M''$ respectively while the red and green solid lines are the fits to $M'$ and $M''$. The insets show the deviation of the fit to $M'$ from the experimentally measured value, in the low $I_{ac}$ regime. Magnetic field dependence of **(g)** $\alpha_0$, (ref. [16], using GR model, neglecting flux creep, (red solid circles) and this analysis (black solid circles)) **(h)** $\frac{U_0}{k_B}$ and **(i)** $J_c$ which are the parameters used for fitting with the Beasley creep model at $T = 2\ K$ and $f = 30\ kHz$.

*NbN:* We show the fits to the amplitude dependence of $M'$ and $M''$ for NbN at different DC magnetic fields in fig. (8 (a) – (f)). The Beasley model of flux creep captures the variation in $M'$ quantitatively, with a small deviation at higher $I_{ac}$. However, in this case at very small $I_{ac}$ ($< 0.7\ mA$), we also observe a significant deviation between the fit and the experimental $M'$, as see from the expanded views in the insets of fig. (8 (a) – (f)). The near linear variation of $M'$ in the low $I_{ac}$ regime could not be fitted with any of the existing models of flux creep and needs to be studied further. The overall $I_{ac}$ variation in $M''$ is captured qualitatively, but with an overestimation, like $a$-MoGe although the low $I_{ac}$ and magnetic fields, the fits and the experimental $M''$ are in very good agreement with each other. The fits to the complete dataset for NbN is provided in Appendix B. Fig. 8 ((g) – (i)) show the magnetic field variation of the fitting parameters: $\alpha_0, U_0,$ and $J_c$ respectively. The values of $J_c$ are of the same order of magnitudes as the ones obtained from earlier studies on similar samples[59,60]. The magnitude of $\alpha_0$ is very close to the one obtained from $\tilde{\lambda}$ calculated from $M$ measured at the lowest $I_{ac}$ using the analysis based on the GR model (eqn. (3)) in ref. [16], since the effect of creep is much less pronounced in NbN than in $a$-MoGe at low $I_{ac}$. The difference in the two values is primarily because we cannot capture the variation in $M'$ at low current using the creep models used here. The magnitude of $U_0$ differs from ref. [16], for the reason similar to $a$-MoGe.

## 5.3 Effect of flux creep on the pinning force constant $\alpha$

We have seen in fig. (7 (g)) that for $a$-MoGe the value of $\alpha$ obtained from analysis of the data at the lowest $I_{ac} \sim 0.05\ mA$, differs significantly from the value obtained from the analysis here that includes the effect of flux creep. This comes from the fact that for very weakly pinned superconductors such as the $a$-MoGe thin film, $J_c$ is small and neglecting the effect of flux creep results in underestimating $\alpha$.

To elucidate this point, we calculate the apparent Labusch parameter $\tilde{\alpha}$ as a function of $I_{ac}$ from $M'$ and $M''$ neglecting the effect of flux creep on the current density profile. For this we first invert the experimentally measured $M'$ and $M''$ using the scheme outlined in ref. [16] to obtain an effective $\tilde{\lambda}$ that is independent of the induced current density. Then using eqn. (3)

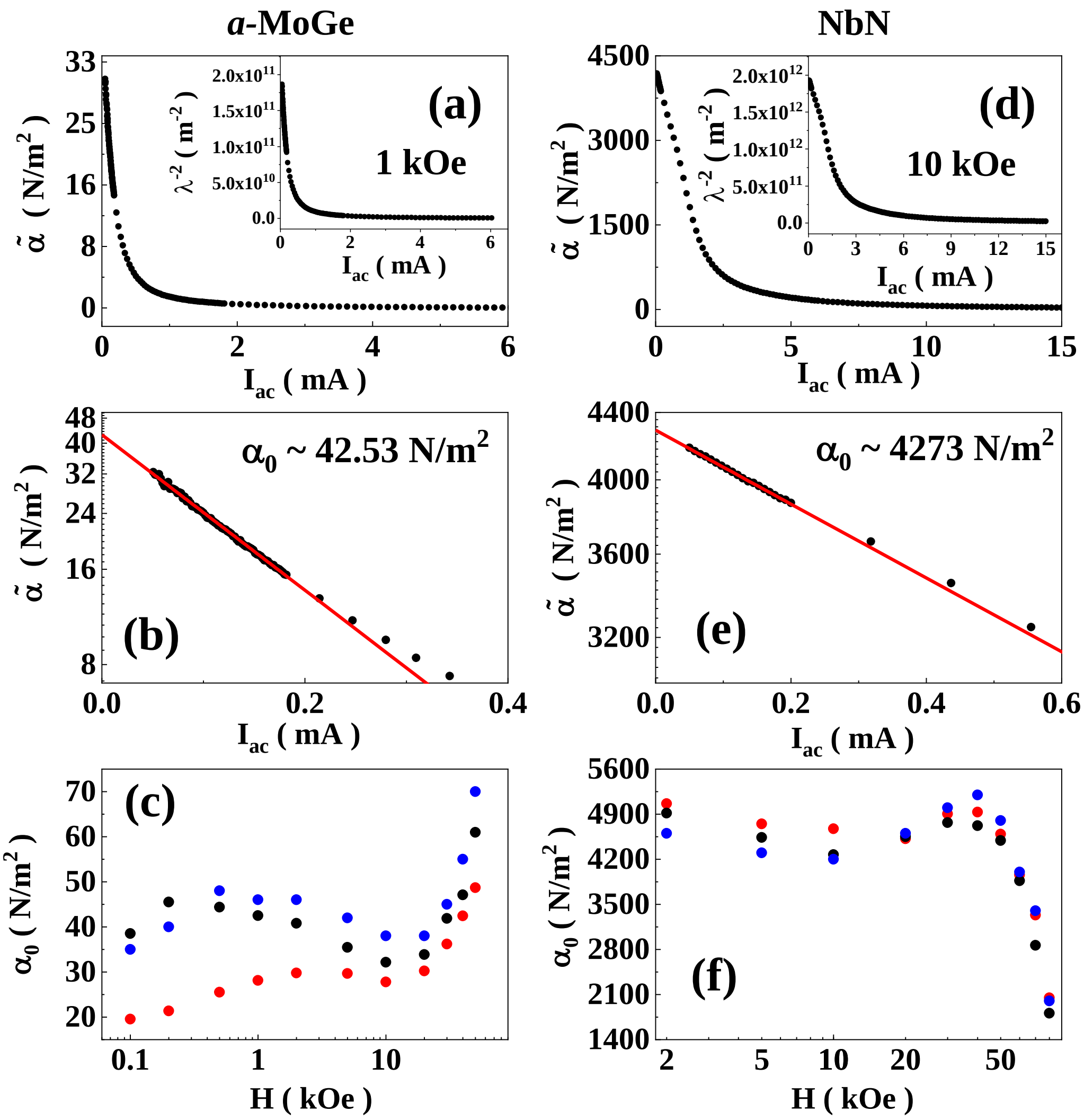

**Figure 9.** Variation of $\tilde{\alpha}$ with $I_{ac}$ for **(a)** $a$-MoGe at 1 $kOe$ and **(d)** NbN at 10 $kOe$ (black solid circles) extracted from $\lambda$ using the GR model, ignoring flux creep (eqn. (3)). Inset shows the $I_{ac}$ variation of $\lambda^{-2}$, calculated by inverting $M'(M'')$ as done in ref. [16]. Expanded view of $\tilde{\alpha}$ vs $I_{ac}$ (black solid circles) in log-linear scale for **(b)** $a$-MoGe and **(e)** NbN in the low $I_{ac}$ regime. The red solid line is the fit to $\tilde{\alpha}$ in the low $I_{ac}$ regime, which is extrapolated for $I_{ac} \approx 0$, to obtain $\alpha_0$. The magnetic field variation of $\alpha_0$ for **(c)** $a$-MoGe and **(f)** NbN respectively. The black solid circles correspond to $\alpha_0$ estimated from the linear extrapolation of $\tilde{\alpha}$ for $I_{ac} \approx 0$. The red solid circles correspond to $\alpha_0$ obtained using the GR model in ref. [16]. The blue solid circles correspond to $\alpha_0$ used as a parameter for fitting $M'(M'')$ $vs$ $I_{ac}$ under the effect of flux creep.

we obtain $\tilde{\alpha}$. Figs. (9 (a)) and (d) show that for the $a$-MoGe and NbN thin films $\tilde{\alpha}$ decreases rapidly with $I_{ac}$. To estimate $\alpha_0$ (defined in the limit $I_{ac} \approx 0$), for we plot $\tilde{\alpha} - I_{ac}$ in log-linear scale and linearly extrapolate the value for $I_{ac} \approx 0$. At low $I_{ac}$, $\tilde{\alpha}$ decreases more rapidly for $a$-MoGe (fig. (9 (b))) than for NbN (fig. (9 (e))). This can be understood from the difference in $J_c$ in the two samples: in $a$-MoGe, which is more weakly pinned than NbN, $J_c$ is two orders of magnitude smaller than NbN and hence the effect of flux creep is stronger. In fig. (9 (c)), for $a$-MoGe, we plot together this extrapolated value of $\tilde{\alpha}$ for $I_{ac} \approx 0$ at different magnetic fields (black solid circles) along with $\alpha_0$ obtained from the full creep analysis (blue solid circles). These two values are very close to each other, whereas the value obtained by neglecting flux creep from the GR analysis is much smaller. In the strongly pinned NbN film (fig. (9 (f))), $\alpha_0$ estimated from extrapolating $I_{ac} \approx 0$ and that obtained using the GR analysis, are very close to each other. This method outlined above provides a simple scheme to obtain $\alpha_0$ without doing the complete analysis involving creep.

Quantitatively, the effect on $\tilde{\alpha}$ when creep is neglected can be understood as follows. Flux creep results in a renormalization of $\alpha$ with $|J_{ac}|$, where $\tilde{\alpha}$ is given by,

$$\tilde{\alpha}(J_{ac}) \sim \alpha(J_{ac}) \times \left( \frac{1-\epsilon}{1+\left(\frac{\epsilon}{\omega\tau_0}\right)^2} \right) \qquad (13)$$

The calculation of this expression follows from eqn. (12) and is given in details in Appendix C. The first factor on the right hand side, $\alpha(J_{ac})$ contains direct effect of $|J_{ac}|$ on $\alpha$ due to shift of the equilibrium position of the vortex in the pinning potential $U$, under a drive (eqn. (10)). There is a second correction that is encapsulated in the second factor. The numerator of the second factor, has $\epsilon$ (eqn. (5 (a))) which is a measure of the attenuation of $U$, through the effect of flux creep (eqn. (7)). For small values of $J_c$, the increase of $\epsilon$ is much more rapid with increasing $|J_{ac}|$ and hence the effect of flux creep is more conspicuous in weakly pinned systems. At low $|J_{ac}|$, $\epsilon$ is $\sim 0.03$ for $a$-MoGe at $1\ kOe$, and $\sim 0.002$ for NbN at $10\ kOe$, implying that in the weakly pinned $a$-MoGe, effect of flux creep manifested through the attenuation of $U$, is stronger than in NbN. However, more significant effect comes from the denominator. The denominator $1+\left(\frac{\epsilon}{\omega\tau_0}\right)^2$, can be taken $\sim 1$ if $\omega \gg \tau_0{}^{-1}$. This happens when the probing frequency is sufficiently high. In our case we are in the low frequency regime (tens of $kHz$). $\tau_0$ (eqn. 5 (b)) which is given by $\eta/\alpha$, is large enough such that $\omega\tau_0 \leq 1$. For $a$-MoGe at $1\ kOe$, $\omega\tau_0$ ranges from $0.038 - 0.004$, with increasing $|J_{ac}|$ such that $\left(\frac{\epsilon}{\omega\tau_0}\right)^2$ $\sim 0.6 - 1700$. In the case of NbN, $\omega\tau_0$ varies from $0.01$ till $0.0007$, which in turn makes $\left(\frac{\epsilon}{\omega\tau_0}\right)^2$ vary from $\sim 0.04$ till $950$ at $10\ kOe$. Therefore, even if $\epsilon \ll 1$, $\tilde{\alpha}$ scales as $\left(\frac{\epsilon}{\omega\tau_0}\right)^{-2}$, which makes $\tilde{\alpha} \ll \alpha$ particularly at high $|J_{ac}|$.

## 6. Conclusion

We have shown that flux creep can give rise to strong nonlinearity in the electrodynamic response in the vortex state of superconducting thin films that extends down to very low ac excitations. The non-linear vortex response, which in our experiment is embedded in the variation of $M$ with $I_{ac}$ can be described by combining the effect of flux creep with the model developed by Coffey and Clem[35]. Our results lend support to the suggestion that at subcritical currents the vortex viscosity is exponentially larger than the Bardeen-Stephen estimate (eqn. (11)) even though its microscopic origin needs to be investigated further. It is important to note that this nonlinearity extends to magnetic fields where the magnetisation vs. magnetic field curve has closed and therefore the superconductor has no apparent irreversibility. Neglecting flux creep when analysing the data can result in underestimating the Labusch parameter, $\alpha$, particularly in weakly pinned thin films, where the effect of creep is more important. In addition, we have also presented a simple extrapolation scheme to obtain the correct value $\alpha$ from the non-linear ac response.

Although our analysis can capture the variation of $M$ vs. $I_{ac}$ within reasonable accuracy, some outstanding questions remain. First, in NbN the weak variation of $M'$ at small $I_{ac}$ could not be captured with any of the existing models of flux creep. Secondly, our analysis overestimates the magnitude $M''$ compared to experiments. These discrepancies point to the limitations of existing theoretical models and need to be addressed in future studies.

## 7. Acknowledgement

This work was financially supported by Department of Atomic Energy, Government of India. We thank Ganesh Jangam for help with magnetisation measurements.

## 8. Author Contribution

SB performed the ac shielding response measurements and analysed the data. Sample growth and basic characterisation was done by SB, SM, VB and JJ. PR conceived the problem and supervised the project. The paper was written by SB and PR. All authors commented on the manuscript.

## 9. Data availability

The datasets analyzed in this work and the corresponding figures can be found in the following https://doi.org/10.5281/zenodo.15030698. For now, the files are restricted to users with access only. However, it can be made accessible upon request.

## Appendix A

### 1) Calculation of $\boldsymbol{J}_{ac}(\boldsymbol{r})$ and $\tilde{\lambda}(\boldsymbol{r})$ considering effect of flux creep

The induced current density in the superconducting film, $J_{ac}(r)$, is obtained by solving the coupled Maxwell's and London equation for the geometry of our system, and is given by the following self-consistent equations,

$$\boldsymbol{A}_{film}(\boldsymbol{r}) = \boldsymbol{A}_{drive}(\boldsymbol{r}) + \frac{\mu_0}{4\pi}\int d^3\boldsymbol{r}'\frac{\boldsymbol{J}_{ac}(\boldsymbol{r}')}{|\boldsymbol{r}-\boldsymbol{r}'|} \quad (14)$$

$$\boldsymbol{A}_{film}(\boldsymbol{r}) = -\mu_0\tilde{\lambda}^2\boldsymbol{J}_{ac}(\boldsymbol{r}) \quad (15)$$

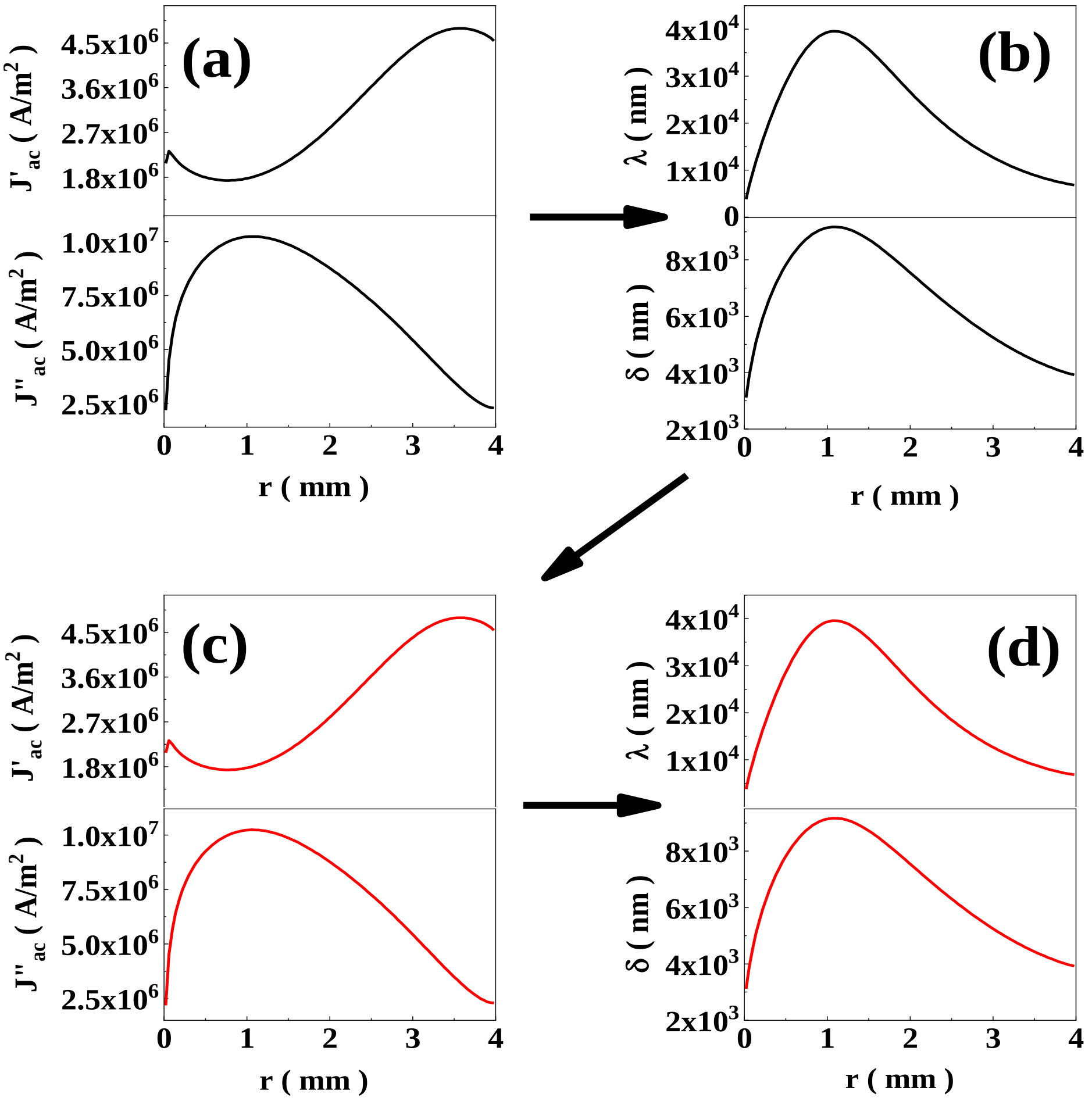

**Figure 10.** Radial variation of $J'_{ac}$, $J''_{ac}$, and $\lambda$, $\delta$ at $I_{ac} = 2\ mA$. The profile of $J'_{ac}$ ($J''_{ac}$) in **(a)** is used to calculate $\lambda$ ($\delta$) in **(b)**, which in turn is used to calculate the profile of $J'_{ac}$ ($J''_{ac}$) in **(c)**. The profile of $\lambda$ ($\delta$) in **(d)** is obtained using $J'_{ac}$ ($J''_{ac}$) in **(c)**. The black and red solid lines correspond to the $1^{st}$ and $2^{nd}$ iterations respectively of the self-consistency check procedure. We have taken $\alpha_0 \sim 46\ Nm^{-2}$, $\frac{U_0}{k_B} \sim 13.2\ K$ and $J_c \sim 4.8 \times 10^7 Am^{-2}$ along with $\eta = \eta_{BS} e^{U_0/k_B T}$ where $\eta_{BS} \sim 1.47 \times 10^{-8}\ Nsm^{-2}$ at $T = 2\ K$ and $B = 1\ kOe$.

where $\boldsymbol{A}_{film}(\boldsymbol{r})$ is the vector potential of the superconducting film. It is a combination of the vector potential associated with the primary coil, $\boldsymbol{A}_{drive}(\boldsymbol{r})$, due to passage of an ac current through it, and the induced current density, $J_{ac}$, on the superconducting film and $\tilde{\lambda}$ is the complex penetration depth. The scheme for numerically solving these equations for a circular geometry and a constant $\tilde{\lambda}$ and obtaining $M'(M'')$, is given in refs. [11] and [12].

When the effect of flux creep is included, additional complications arise when $\tilde{\lambda}$ becomes a function of the induced current density $J_{ac}$ which is function of the radial distance from the center of the film $r$ . The relation between $J_{ac}(r)$ and $\tilde{\lambda}$, can be calculated using eqns. (7), (10), (11) and (12). To obtain a self-consistent solution between $J_{ac}(r)$ and $\tilde{\lambda}$, we adopt an iterative procedure. In the first step, assume constant values for $\alpha_0$, $\frac{U_0}{k_B}$ and $\eta \sim \eta_{BS} e^{\frac{U_0}{k_B T}}$ and calculate $J_{ac}(r)$ using eqns. (14) and (15). In the second step, we use the calculated $J_{ac}(r)$ in the first step and calculate $\tilde{\lambda}(r)$ using eqns. (7), (10), (11) and (12) and calculate $J_{ac}(r)$ using this modified $\tilde{\lambda}(r)$. This procedure is then repeated till a convergence is reached.

While this basic procedure leads to a convergence at low values of $I_{ac}$, as $I_{ac}$ is increased, this iteration fails to converge. We therefore modify the protocol in the following way. From the 7$^{th}$ iteration step we take the average of the $\tilde{\lambda}(r)$ in the last 6 iterations to calculate $J_{ac}(r)$. This protocol converges up to the maximum $I_{ac}$ in about 300 iterations.

To cross-check the self-consistency of the procedure, we use the final $J_{ac}(r)$ obtained after convergence and use it to directly compute $\tilde{\lambda}(r)$, using eqns. (7), (10), (11) and (12). Next, using the computed value of $\tilde{\lambda}(r)$, we recalculate $J_{ac}(r)$ using eqns. (14) and (15) and verify that we get the same result. This is shown in fig. (10) for *a*-MoGe at $B = 1\, kOe$ at $I_{ac} = 2\, mA$.

## Appendix B

**Complete dataset of *a*-MoGe and NbN for the fits to the electromagnetic shielding response as a function of $I_{ac}$ at different DC magnetic fields**

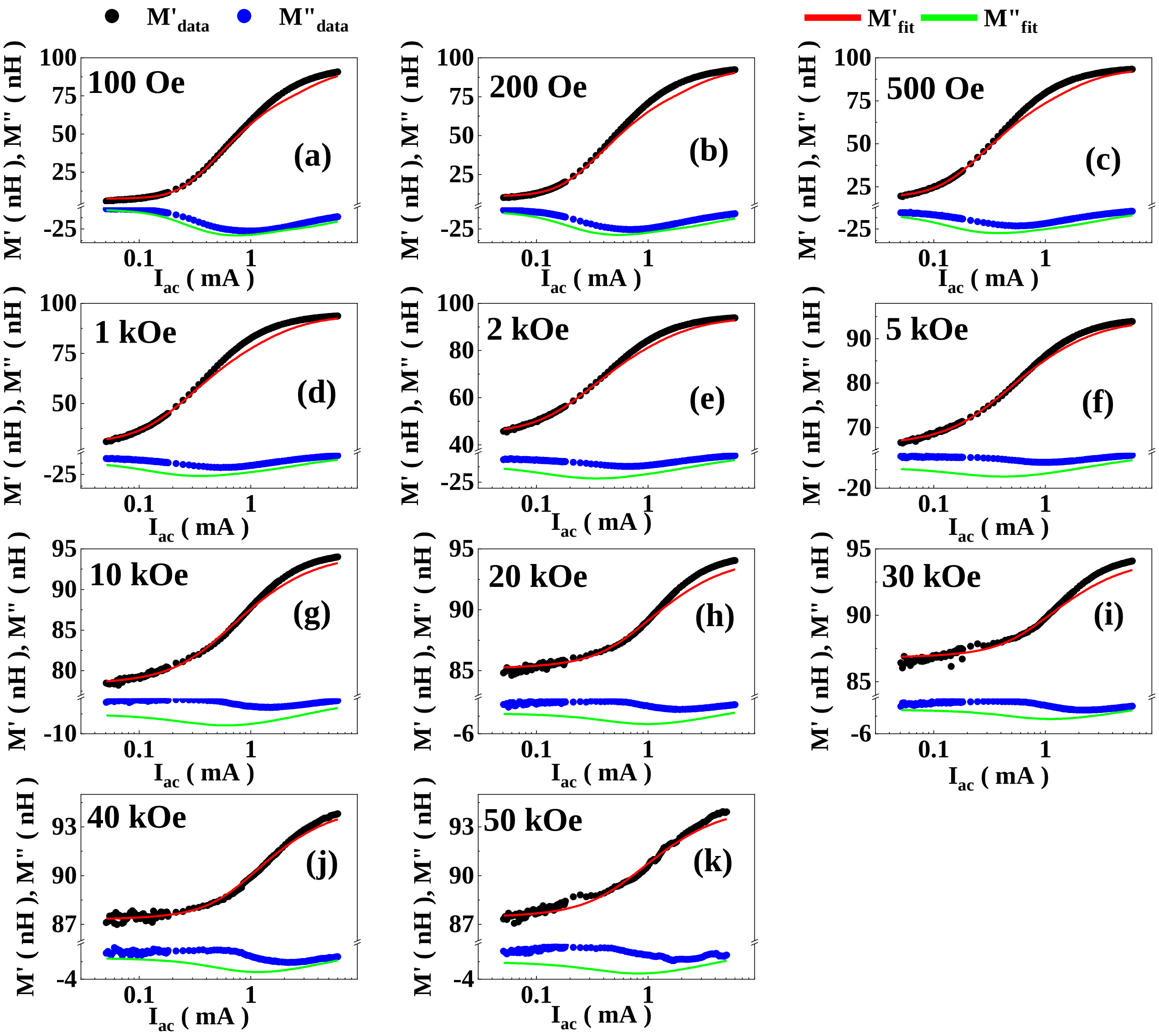

**Figure 11. (a) – (k)** $I_{ac}$ dependence of $M'$ and $M''$ of 20 $nm$ *a*-MoGe at different applied DC magnetic fields. The black and blue solid circles are the experimentally measured $M'$ and $M''$ respectively while the red and green solid lines are the fits to $M'$ and $M''$. The measurements are done at $T = 2\ K$ and $f = 30\ kHz$.

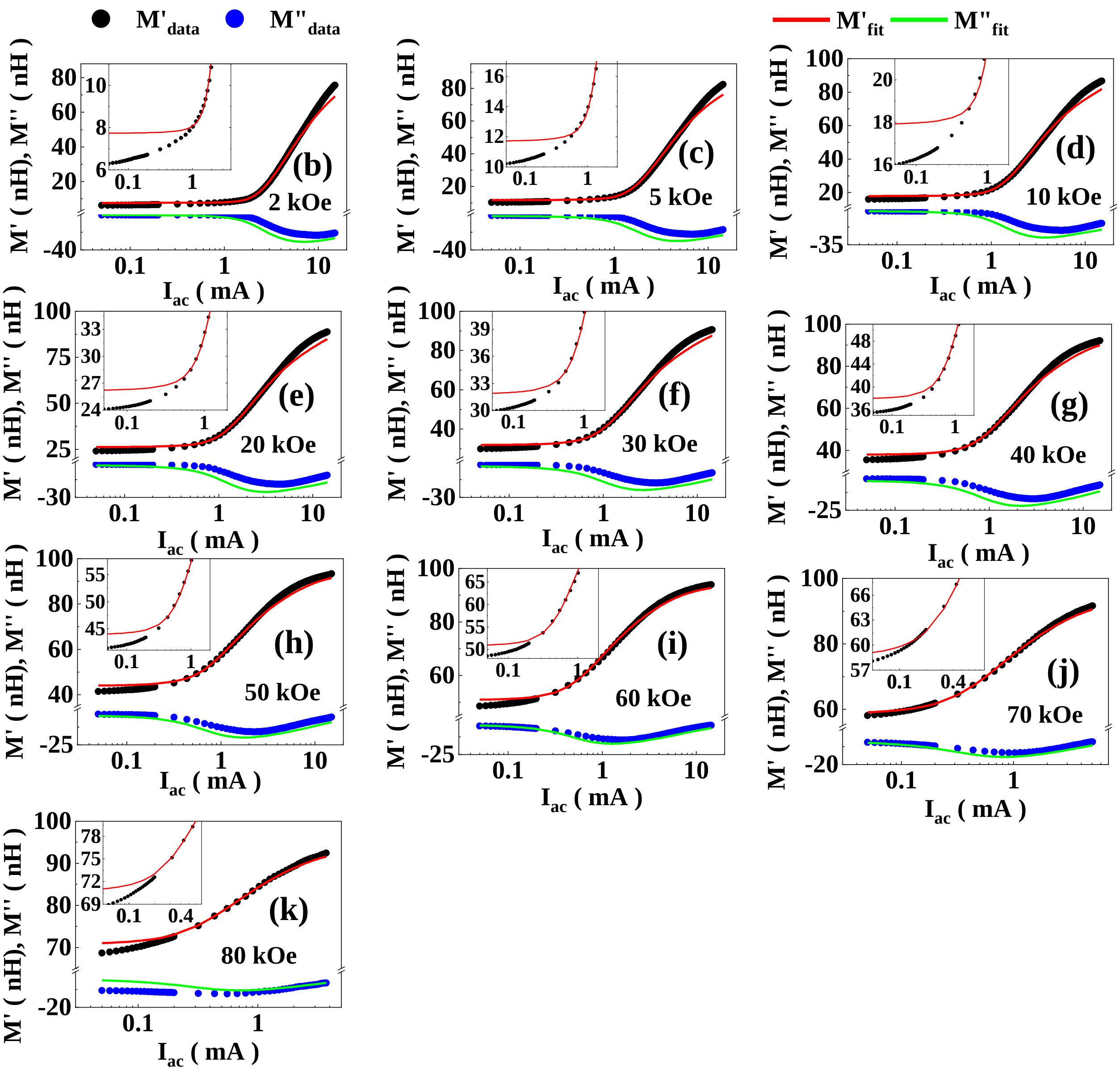

**Figure 12. (a) – (k)** $I_{ac}$ dependence of $M'$ and $M''$ of 5 $nm$ NbN at different applied DC magnetic fields. The black and blue solid circles are the experimentally measured $M'$ and $M''$ respectively while the red and green solid lines are the fits to $M'$ and $M''$. The insets (X-axis in $mA$ and Y-axis in $nH$) show the deviation of the fit to $M'$ from the experimental data at low $I_{ac}$. The measurements are done at $T = 2\ K$ and $f = 30\ kHz$.

## Appendix C

### Calculation of $\alpha_L$ and $\eta$ under the influence of flux creep

We have the CC eqn. (12) which is : $\tilde{\lambda}^{-2} = \lambda^{-2} + i\delta^{-2} = \left(\lambda_L^2 - i\frac{\rho_{ff}}{\mu_0\omega}\frac{\epsilon+i\omega\tau}{1+i\omega\tau}\right)^{-1}$ where $\epsilon = \frac{1}{I_0^2(\nu)}$ and $\tau = \tau_0\frac{I_0^2(\nu)-1}{I_1(\nu)I_0(\nu)}$ (eqn. (5)) and $\tau_0 = \eta/\alpha$. In our analysis, $\frac{I_0^2(\nu)-1}{I_1(\nu)I_0(\nu)} \sim 1.15 - 1.18$ and hence for simplicity in the calculations to follow, we assume $\tau \sim \tau_0$. Eqn. (12) can be simplified as follows:

$$\tilde{\lambda}^2 \sim \lambda_L^2 - i\frac{\rho_{ff}}{\mu_0\omega}\frac{\epsilon + i\omega\tau_0}{1 + i\omega\tau_0}$$

$$\Rightarrow \tilde{\lambda}^2 - \lambda_L^2 \sim \frac{\rho_{ff}}{\mu_0\omega}\frac{\omega\tau_0 - i\epsilon}{1 + i\omega\tau_0}$$

$$\Rightarrow \tilde{\lambda}^2 - \lambda_L^2 \sim \frac{B\phi_0}{\mu_0\omega\eta}\frac{\omega\tau_0 - i\epsilon}{1 + i\omega\tau_0}$$

$$\Rightarrow \left(\tilde{\lambda}^2 - \lambda_L^2\right)^{-1} \sim \frac{\mu_0\omega\eta}{B\phi_0}(1 + i\omega\tau_0)(\omega\tau_0 - i\epsilon)^{-1}$$

$$\Rightarrow \left(\tilde{\lambda}^2 - \lambda_L^2\right)^{-1} \sim \frac{\mu_0\eta}{B\phi_0\tau_0}(1 + i\omega\tau_0)\left(1 - \frac{i\epsilon}{\omega\tau_0}\right)^{-1}$$

$$\Rightarrow \left(\tilde{\lambda}^2 - \lambda_L^2\right)^{-1} \sim \frac{\mu_0\alpha}{B\phi_0}(1 + i\omega\tau_0)\left(1 - \frac{i\epsilon}{\omega\tau_0}\right)^{-1}$$

$$\Rightarrow \left(\tilde{\lambda}^2 - \lambda_L^2\right)^{-1} \sim \frac{\mu_0}{B\phi_0}(\alpha + i\omega\eta)\left(1 - \frac{i\epsilon}{\omega\tau_0}\right)^{-1}$$

$$\Rightarrow \left(\tilde{\lambda}^2 - \lambda_L^2\right)^{-1} \sim \frac{\mu_0}{B\phi_0}\frac{(\alpha + i\omega\eta)}{\left(1 - \frac{i\epsilon}{\omega\tau_0}\right)}$$

$$\Rightarrow \left(\tilde{\lambda}^2 - \lambda_L^2\right)^{-1} \sim \frac{\mu_0}{B\phi_0}\frac{(\alpha + i\omega\eta)}{1 + \left(\frac{\epsilon}{\omega\tau_0}\right)^2}\left(1 + \frac{i\epsilon}{\omega\tau_0}\right)$$

$$\Rightarrow \left(\tilde{\lambda}^2 - \lambda_L^2\right)^{-1} \sim \frac{\mu_0}{B\phi_0} \frac{1}{1 + \left(\frac{\epsilon}{\omega\tau_0}\right)^2} \left(\alpha + \frac{i\epsilon\alpha^2}{\omega\eta} + i\omega\eta - \alpha\epsilon\right)$$

$$\Rightarrow \left(\tilde{\lambda}^2 - \lambda_L^2\right)^{-1} \sim \frac{\mu_0}{B\phi_0} \frac{1}{1 + \left(\frac{\epsilon}{\omega\tau_0}\right)^2} \left(\alpha(1 - \epsilon) + i\omega\eta\left(1 + \epsilon\left(\frac{1}{\omega\tau_0}\right)^2\right)\right)$$

$$\Rightarrow \left(\tilde{\lambda}^2 - \lambda_L^2\right)^{-1} \sim \frac{\mu_0}{B\phi_0}(\tilde{\alpha} + i\omega\tilde{\eta}) \qquad (16)$$

where $\tilde{\alpha}$ and $\tilde{\eta}$ are the pinning force constant and vortex viscosity, renormalized due to the effect of flux creep. These quantities are related to their creep free counterparts as:

$$\tilde{\alpha} \sim \alpha \frac{(1-\epsilon)}{1+\left(\frac{\epsilon}{\omega\tau_0}\right)^2} \qquad (17\text{ (a)})$$

$$\tilde{\eta} \sim \eta \frac{\left(1+\epsilon\left(\frac{1}{\omega\tau_0}\right)^2\right)}{1+\left(\frac{\epsilon}{\omega\tau_0}\right)^2} \qquad (17\text{ (b)})$$